\documentclass[aps,prl,superscriptaddress,notitlepage,floatfix,twocolumn]{revtex4-1}
\usepackage{amsmath}
\usepackage{graphicx}   
\usepackage{dcolumn}    
\usepackage{bm}         
\usepackage{amssymb}    
\usepackage{float}
\usepackage{caption}
\usepackage{subcaption}
\usepackage{verbatim}
\usepackage{color}
\usepackage[usenames,dvipsnames,svgnames,table]{xcolor}
\usepackage[colorlinks=true,
            linkcolor=red,
            urlcolor=blue,
            citecolor=green,
            bookmarks=true,
            bookmarksnumbered=true,
            breaklinks=true,
            pdfpagemode=Fullscreen,
            pdfstartview=FitBH]{hyperref}

\definecolor{gesfred}{rgb}{1,0,0}

\definecolor{gesfblue}{rgb}{0.08,0.42,0.76}

\definecolor{gesfgreen}{rgb}{0.0,0.65,0.0}

\begin{document}


\hfill KEK-TH-1742
\vspace{10mm}

\title{Unify Residual $\mathbb Z^{23}_2 \otimes \mathbb Z^{12}_2$ Symmetries and Quark-Lepton Complementarity}
\author{Shao-Feng Ge}
\email{gesf02@gmail.com}
\affiliation{KEK Theory Center, Tsukuba 305-0801, Japan}
\date{\today}

\begin{abstract}
The residual $\mathbb Z^{23}_2$ ($\mathbb Z^{\mu\tau}_2$) and 
$\mathbb Z^{12}_2$ ($\mathbb Z^s_2$ or $\overline{\mathbb Z}^s_2$) symmetries 
can not only apply to the lepton sector, but also explain the mainly $1$--$2$ 
mixing in the quark sector.
With both up and down quarks subject to the same $\mathbb Z^{23}_2$ 
symmetry, which interchanges the second and third generations, 
the $1$--$3$ and $2$--$3$ mixing angles in CKM vanish while the Cabibbo angle 
is dictated by two residual $\mathbb Z^{12}_2$ symmetries. It turns out
that the quark-lepton complementarity conditions are natural consequences of 
unifying residual $\mathbb Z^{23}_2 \otimes \mathbb Z^{12}_2$ symmetries in 
both lepton and quark sectors. 
\end{abstract}

\maketitle

\noindent {\bf Introduction}--
The idea of residual symmetry \cite{residual1, residual2} in the lepton sector is based on the fact that 
fermion mixing comes from diagonalization of the 
corresponding mass matrix. If there is any symmetry that 
dictates the mixing pattern after fermion becomes massive, it has to be residual symmetry that survives 
symmetry breaking. Although fundamental lagrangian can be subject to
some full flavor symmetries, the mixing matrix is not necessarily determined
by them as full symmetries have to be broken for chiral fermions to acquire mass terms. 
There are other factors, such as VEV dictated by scalar coupling constants whose 
magnitude is not under the control of flavor symmetry. Yukawa coupling constant,
especially its magnitude, is another instance. On the other hand,
full flavor symmetry has to apply on $SU(2)_L$ doublet as a whole, in other words,
equally for its two component fermions. If full flavor symmetry is not
broken, the charged lepton and neutrino mass matrices are subject to a same
symmetry, leading to a same mixing matrix. The physical mixing matrix becomes
trivial in the absence of mismatch between the lepton and neutrino mixing matrices.

We can find analogies in electro-weak symmetry breaking (EWSB) for the above two 
arguments.  First, the gauge bosons are predicted by gauge symmetry but the mixing
between them is determined by the gauge coupling constants whose magnitude cannot 
be determined by the gauge symmetry. Instead, the gauge boson masses and the weak 
mixing angle are correlated by the so-called custodial symmetry \cite{custodial}. Second, 
if gauge symmetry is not broken, no mixing between gauge bosons can be possible.
In some sense, both residual and custodial symmetries serve as effective theories
of the corresponding fundamental full symmetries at low energy.

Phenomenologically speaking, residual symmetry such as $\mathbb Z^s_2$ and 
$\overline{\mathbb Z}^s_2$ can predict unique correlations between mixing angles 
and the Dirac {\tt CP} phase \cite{residual1,residual2,Hanlon:2013ska}. Note that, only observable parameters are involved in 
this correlation which is also similar to the prediction of the custodial 
symmetry in EWSB between gauge boson masses and mixing. 
They can serve as a robust test of the symmetries behind them
and can also predict the less precisely measured parameters with the others.
Reversely, residual symmetries can be used to reconstruct the full flavor 
symmetry \cite{reconstruction} in a bottom-up approach.

Since residual symmetry has been quite successful in the lepton sector \cite{review}, 
it can be a natural generalization to the quark sector. 
Unlike the PMNS matrix \cite{PMNS} of lepton mixing where all three families 
are mixed with sizable amount, the quark mixing in the CKM matrix
\cite{CKM} happens mainly between the first two families. 
Of the two residual symmetries, $\mathbb Z^{\mu\tau}_2$ and $\mathbb Z^s_2$
($\overline{\mathbb Z}^s_2$), the first determines the 1--3 and 2--3 mixing 
angles while the second constrains the 1--2 mixing angle. If 
$\mathbb Z^{\mu\tau}_2$ can serve as residual symmetry for both up and 
down quarks, we denote it as $\mathbb Z^{23}_2$ hereafter for generality since it 
interchanges the second and third generations, same 1--3 and 2--3 mixing
angles appear in the up and down quark mixing matrices. In other words, 
there is no mismatch between 1--3 and 2--3 quark mixings, leading to tiny
mixing between the third generation and the other two in the CKM matrix
\cite{23}.
This success leads us to think about the possibility of generalizing and
unifying the two residual $\mathbb Z_2$ symmetries in both lepton and quark 
sectors. 

Another relevant
ingredient is the so-called quark-lepton complementarity 
\cite{complementarity} that relates the quark and neutrino
mixing angles at the leading order. If the residual symmetries
in both sectors can be unified, the quark-lepton complementarity becomes
a natural consequence. This shows the power of residual $\mathbb Z_2$
symmetries.

%
%

\noindent {\bf Residual $\bf \boldsymbol{\mathbb Z}^{\bf 23}_2$ Symmetry}--
The oscillation-relevant part of the neutrino mixing matrix is the so-called 
PMNS matrix \cite{PMNS}. In the standard parametrization \cite{pdg12}, 
it takes the form as,
$V_{\rm PMNS} = $
\begin{equation*}
\hspace{-1mm}
\left\lgroup
\begin{matrix}
  c^\nu_{12} c^\nu_{13} 
& s^\nu_{12} c^\nu_{13} 
&
\hspace{-2mm}
 s^\nu_{13} e^{- i \delta^\nu_D} 
\\
\hspace{-1mm}
- s^\nu_{12} c^\nu_{23} \hspace{-1mm} - \hspace{-1mm} c^\nu_{12} s^\nu_{23} s^\nu_{13} e^{i \delta^\nu_D}
&
\hspace{-2mm}
 c^\nu_{12} c^\nu_{23} \hspace{-1mm} - \hspace{-1mm} s^\nu_{12} s^\nu_{23} s^\nu_{13} e^{i \delta^\nu_D}
&
\hspace{-2mm}
 s^\nu_{23} c^\nu_{13} 
\\
\hspace{-1mm}
  s^\nu_{12} s^\nu_{23} \hspace{-1mm} - \hspace{-1mm} c^\nu_{12} c^\nu_{23} s^\nu_{13} e^{i \delta^\nu_D}
&
\hspace{-2mm}
-c^\nu_{12} s^\nu_{23} \hspace{-1mm} - \hspace{-1mm} s^\nu_{12} c^\nu_{23} s^\nu_{13} e^{i \delta^\nu_D}
&
\hspace{-2mm}
 c^\nu_{23} c^\nu_{13} 
\end{matrix}
\right\rgroup 
\hspace{-1mm}
,
\end{equation*}
where $(c^\nu_{ij}, s^\nu_{ij}) \equiv (\cos \theta^\nu_{ij}, \sin \theta^\nu_{ij})$ and $\delta^\nu_D$ is the Dirac {\tt CP} phase. To avoid confusion, we have
added a subscript $\nu$ to all the mixing parameters. In the following 
discussions, we will generalize it to both up and down quarks by replacing
the subscript $\nu$ with $u$ and $d$, respectively.

Note that, vanishing reactor mixing angle ($\theta^\nu_{13} = 0^\circ$) and 
maximal atmospheric mixing angle ($\theta^\nu_{23} = 45^\circ$) is a good 
zeroth-order approximation. In the diagonal basis of charged leptons, this is 
a natural consequence of the so-called $\mu$--$\tau$ symmetry \cite{mutau}, 
denoted as 
$\mathbb Z^{\mu\tau}_2$. The nonzero $\theta^\nu_{13}$ and the deviation of
$\theta^\nu_{23}$ from being maximal indicates the extent of
$\mu$--$\tau$ symmetry breaking. 

The $\mu$ and $\tau$ flavors interchange under the $\mathbb Z^{\mu\tau}_2$ 
representation, 
\begin{equation}
  G(\mathbb Z^{\mu\tau}_2)
\equiv
  G_{23}
=
\left\lgroup
\begin{matrix}
  1 & 0 & 0 \\
  0 & 0 &-1 \\
  0 &-1 & 0
\end{matrix}
\right\rgroup \,.
\end{equation}
If the neutrino mass matrix is invariant under $G_{23}$,
the number of independent matrix elements can be reduced and the mixing 
parameters are constrained. There is a direct relation between horizontal
symmetry transformation representation $G_i$ and the mixing matrix $V_\nu$
\cite{horizontal},
\begin{equation}
  G_i = V_\nu d_i V^\dagger_\nu \,,
\label{eq:GVdV}
\end{equation}
where $d_i$ ($i = 1, 2, 3$) is a diagonal matrix,
\begin{equation}
  d_1
\equiv 
\left\lgroup
\begin{matrix}
- 1 \\
& 1 \\
& & 1
\end{matrix}
\right\rgroup \,,
\quad
  d_2
\equiv 
\left\lgroup
\begin{matrix}
  1 \\
&-1 \\
& & 1
\end{matrix}
\right\rgroup \,,
\label{eq:di}
\end{equation}
and $d_3 \equiv - d_1 d_2$. The relation (\ref{eq:GVdV}) is essentially 
the reconstruction formula of $G_i$ and we call $d_i$ the reconstruction
{\it kernel} \cite{residual1}. 
When viewed in the reversed way, the mixing matrix can 
diagonalize not only the mass matrix but also the horizontal symmetry 
transformation representation. For $G_{23} = G_3 (\theta_{13} = 0, \theta_{23} = 45^\circ)$, it is enough to 
have a maximal 2--3 mixing to diagonalize it into $d_3 = \mbox{diag}(1,1,-1)$. 
Further, there is degeneracy between the first two eigenvalues in $d_3$, 
indicating a free 1--2 mixing. On the other hand, 1--3 mixing is not allowed.
To be exact, the mixing matrix reads \cite{residual1},
\begin{equation}
  V_\nu
=
\left\lgroup
\begin{matrix}
  c^\nu_{12} e^{i \alpha^\nu_1} & s^\nu_{12} e^{i \alpha^\nu_1} & 0 \\
  \frac {-1} {\sqrt 2} s^\nu_{12} & \frac 1 {\sqrt 2} c^\nu_{12} & \frac 1 {\sqrt 2} \\
  \frac 1 {\sqrt 2} s^\nu_{12} & \frac {-1} {\sqrt 2} c^\nu_{12} & \frac 1 {\sqrt 2} \\
\end{matrix}
\right\rgroup
\mathcal Q_\nu \,,
\label{eq:Vnu}
\end{equation}
where 
$Q_\nu \equiv \mbox{diag}(e^{i \phi^\nu_1}, e^{i \phi^\nu_2}, e^{i \phi^\nu_3})$ 
is a diagonal rephasing matrix containing three Majorana (unphysical) {\tt CP} 
phases for Majorana (Dirac) fermions. Note that there is a counterpart of
$\mathcal Q_\nu$, denoted as $\mathcal P_\nu \equiv \mbox{diag}(e^{i \alpha^\nu_1}, e^{i \alpha^\nu_2}, e^{i \alpha^\nu_3})$ which is a purely unphysical rephasing
matrix, to the left of $V_{\rm PMNS}$. 
These unphysical phases are 
constrained to be $e^{i \alpha^\nu_2} = e^{i \alpha^\nu_3}$ 
and can be rotated away, leaving only a free $e^{i \alpha^\nu_1}$ behind.
Note that the notation in this paper is different from the one adopted in
\cite{residual1,residual2} by a minus sign associated with all the mixing angles
and the Dirac {\tt CP} phase.

The horizontal symmetry with real {\it kernel} applies to not only Majorana 
but also Dirac fermions \cite{residual1}. This makes it possible 
to generalize the $\mathbb Z^{\mu \tau}_2$ symmetry in the lepton sector 
to $\mathbb Z^{23}_2$ which is applicable in both lepton and quark sectors.
With both up and down quarks subject to the same residual 
$\mathbb Z^{23}_2$, 
their mixing matrices $V_u$ and $V_d$ are constrained as (\ref{eq:Vnu}),
\begin{subequations}
\begin{eqnarray}
  V_u
& = &
\left\lgroup
\begin{matrix}
  c^u_{12} e^{i \alpha^u_1} & s^u_{12} e^{i \alpha^u_1} & 0 \\
  \frac {-1} {\sqrt 2} s^u_{12} & \frac 1 {\sqrt 2} c^u_{12} & \frac 1 {\sqrt 2} \\
  \frac 1 {\sqrt 2} s^u_{12} & \frac {-1} {\sqrt 2} c^u_{12} & \frac 1 {\sqrt 2} \\
\end{matrix}
\right\rgroup
\mathcal Q_u,
\\
  V_d
& = &
\left\lgroup
\begin{matrix}
  c^d_{12} e^{i \alpha^d_1} & s^d_{12} e^{i \alpha^d_1} & 0 \\
  \frac {-1} {\sqrt 2} s^d_{12} & \frac 1 {\sqrt 2} c^d_{12} & \frac 1 {\sqrt 2} \\
  \frac 1 {\sqrt 2} s^d_{12} & \frac {-1} {\sqrt 2} c^d_{12} & \frac 1 {\sqrt 2} \\
\end{matrix}
\right\rgroup
\mathcal Q_d.
\end{eqnarray}
\end{subequations}
Combining $V_u$ and $V_d$, the CKM matrix $V_{\rm CKM} \equiv V^\dagger_u V_d$ 
reads,
\begin{equation}
\hspace{-2mm}
  V_{\rm CKM}
\hspace{-1mm} = \hspace{-1mm}
\left\lgroup
\begin{matrix}
  c_u c_d + s_u s_d e^{i \delta_\alpha}
& c_u s_d - s_u c_d e^{i \delta_\alpha}
& 0 \\
  s_u c_d - c_u s_d e^{i \delta_\alpha}
& s_u s_d + c_u c_d e^{i \delta_\alpha} 
& 0 \\
  0 & 0 & 1
\end{matrix}
\right\rgroup,
\label{eq:CKM}
\end{equation}
where $(c_q, s_q) \equiv (\cos \theta^q_{12}, \sin \theta^q_{12})$ with $q$ 
being $u$ or $d$ for up and down quarks, respectively. Note that the rephasing matrices
$\mathcal Q_u$ and $\mathcal Q_d$, which do not have any physical parameters, 
have been removed for simplicity. As there is no intrinsic {\tt CP} phase in two 
dimensional rotation, the phase $\delta_\alpha \equiv \alpha^u_1 - \alpha^d_1$ can be rotated out, reducing (\ref{eq:CKM}) 
into a pure $1$--$2$ rotation with Cabibbo angle,
\begin{eqnarray}
  \cos^2 \theta_C
=
  s^2_d s^2_u
+ 2 c_d s_d c_u s_u \cos \delta_\alpha
+ c^2_d c^2_u.
\label{eq:combination}
\end{eqnarray}
We can see that $\theta^q_{13} = \theta^q_{23} = 0^\circ$ in the CKM matrix 
is a direct consequence of applying the residual $\mathbb Z^{23}_2$ symmetry
to both up and down quarks simultaneously. With vanishing $\delta_\alpha$, which
is a reasonable situation for zeroth-order mass matrix being real,
(\ref{eq:combination}) reduces to a much simpler form,
\begin{equation}
  \theta_C 
=
  \theta^d_{12} - \theta^u_{12} \,.
\label{eq:thetaC}
\end{equation}
The tiny mixing between the third family
and the other two can be treated as small perturbations.

Note that, the two complementarity conditions \cite{complementarity}, 
\begin{equation}
  \theta^\nu_{13} = \theta^q_{13} = 0^\circ \,,
\qquad
  \theta^\nu_{23} + \theta^q_{23} = 45^\circ \,,
\label{eq:complementarity1}
\end{equation} 
have already been obtained
with a simple $\mathbb Z^{23}_2$ symmetry. It applies to the neutrino
and quark sectors while the charged lepton sector is in the diagonal basis
with the help of two extra residual $\mathbb Z_2$ symmetries
represented by $d_1$ and $d_2$ in (\ref{eq:di}). In total, we need
a unified group generated by $G_{23}$, $d_1$, and $d_2$. Since $d_1$ 
can commute with $G_{23}$ as well as $d_2$ and the fact that the order of all
three is 2, namely $G^2_{23} = d^2_1 = d^2_2 = I$, any group element can be expressed as a product $(G_{23} d_2)^i G^j_{23} d^k_1$ where $i = 0, 1, 2, 3$ and $j,k = 0, 1$. The unified group is finite.\\

\noindent {\bf Residual $\bf \mathbb Z^{\bf 12}_2$ Symmetry}--
Based on $\mathbb Z^{23}_2$, the remaining 1--2 mixing angle can be 
determined by residual $\mathbb Z^{12}_2$ symmetry \cite{residual1}
whose generator is $G_1$. Since the 1--3 and 2--3 mixing angles have been
constrained to be $0^\circ$ and $45^\circ$, respectively, for neutrino
and quarks, $G_1$ has only one free parameter, the 1--2 mixing angle,
\begin{equation}
  G^f_{12}
\equiv
\left\lgroup
\begin{matrix}
  - \cos 2 \theta^f_{12} 
& \frac {+ 1}{\sqrt 2} \sin 2 \theta^f_{12} 
& \frac {- 1}{\sqrt 2} \sin 2 \theta^f_{12} 
\\
  \frac {+ 1}{\sqrt 2} \sin 2 \theta^f_{12} 
& \cos^2 \theta^f_{12} 
& \sin^2 \theta^f_{12} 
\\
  \frac {- 1}{\sqrt 2} \sin 2 \theta^f_{12} 
& \sin^2 \theta^f_{12} 
& \cos^2 \theta^f_{12} 
\end{matrix}
\right\rgroup \,,
\label{eq:G12}
\end{equation}
where $f = \nu, u, d$. By choosing $G^f_{12}$ properly, we can determine
the three 1--2 mixing angles, $\theta^\nu_{12}$, $\theta^u_{12}$, and 
$\theta^d_{12}$. We will show whether they can form a finite group.

Let us first concentrate on the quark sector. There are three independent
generators, $G_{23}$, $G^u_{12}$, and $G^d_{12}$. Since $G_{23}$ commutes
with $G^f_{12}$, and the order of them is $2$, 
any group element can be expressed as
\begin{eqnarray}
  (G^d_{12})^{i'} (G^u_{12} G^d_{12})^i (G^u_{12})^{i''} (G^q_{23})^k,
\label{eq:elementa}
\end{eqnarray}
where $i$ is a non-negative integer while $i'$, $i''$, and $k$
can be either $0$ or $1$. Whether the group is finite or not depends on 
whether the combination $(G^u_{12} G^d_{12})$
has a finite order or not. It can be explicitly shown that,
\begin{equation}
  (G^{f_1}_{12} G^{f_2}_{12})
=
  G^{f_1 - f_2}_{12} d_1 \,,
\label{eq:G12product}
\end{equation}
where $G^{f_1 - f_2}_{12}$ is an abbreviation for the functional form 
(\ref{eq:G12}) with $\theta^f_{12}$ replaced by 
$\theta^{f_1}_{12} - \theta^{f_2}_{12}$. In addition, we can see that,
\begin{equation}
  d_1 G^f_{12} 
=
  G^{-f}_{12} d_1 \,,
\label{eq:dG12}
\end{equation}
which leads to $(G^u_{12} G^d_{12})^2 = G^{2 (u-d)}_{12} d_1$.
Here, $G^{-f}_{12}$ stands for the functional form 
(\ref{eq:G12}) with $-\theta^f_{12}$ in the place of $\theta^f_{12}$.
When (\ref{eq:G12product}) and (\ref{eq:dG12}) are applied repeatedly, 
(\ref{eq:elementa}) turns into,
\begin{eqnarray}
  (G^d_{12})^{i'} G^{i (u - d)}_{12} d^n_1 (G^u_{12})^{i''} (G^q_{23})^k \,,
\label{eq:elementb}
\end{eqnarray}
where $n = 0, 1$. We can see that whether the group is finite or not 
depends on whether the Cabibbo angle is a rational proportion of $2 \pi$ or not. 
Turning the way 
around, if the residual $\mathbb Z^{23}_2$ and $\mathbb Z^{12}_2$ symmetries
are the real cause behind the zeroth-order CKM matrix, the Cabibbo angle 
$\theta_C$ is predicted to be a rational proportion of $2 \pi$.\\

\noindent {\bf Unify Residual $\mathbb Z^{23}_2$ and $\mathbb Z^{12}_2$ Symmetries}-- It has been proved above that the first two quark-lepton complementarity
conditions (\ref{eq:complementarity1}) come from the residual $\mathbb Z^{23}_2$
symmetry. To account for the remaining one,
\begin{equation}
  \theta^\nu_{12} + \theta_C
=
  \theta^\nu_{12} + \theta^d_{12} - \theta^u_{12} 
=
  45^\circ \,,
\label{eq:complementarity2}
\end{equation}
it is necessary to involve $\mathbb Z^{12}_2$. The situation can be slightly
complicated since we have three extra $G^f_{12}$ generators with parameters 
satisfying (\ref{eq:complementarity2}), namely $G^\nu_{12}$, $G^u_{12}$, and
$G^d_{12}$, in addition to $G_{23}$.

Since $G_{23}$ can commute with all others, any group element can be expressed 
as,
\begin{equation}
  \{ G^\nu_{12}, G^u_{12}, G^d_{12} \} G^m_{23} \,,
\label{eq:full0}
\end{equation}
where $m = 0,1$ and $\{ \cdots \}$ stands for an arbitrary product of the 
generators inside it. Then we can use (\ref{eq:G12product}) and (\ref{eq:dG12})
repeatedly to reduce the first factor in (\ref{eq:full0}) which becomes,
\begin{equation}
  G^{i \times \nu + j \times u + k \times d}_{12} G^m_{23} d^n_1 \,,
\label{eq:full1}
\end{equation}
where $i$, $j$, and $k$ are integers while $n = 0, 1$. 
For the unified group to be finite, the first factor
in (\ref{eq:full1}) has to be cut off somewhere. This means that the three
1--2 mixing angles, $\theta^\nu_{12}$, $\theta^u_{12}$, and $\theta^d_{12}$,
have to be rational proportions of $2 \pi$ 
and correlated according to (\ref{eq:complementarity2}).
This is possible and can find many realizations. As long as the orders of 
$G^\nu_{12}$, $G^u_{12}$, and $G^d_{12}$ are large enough, we can predict 
1--2 mixing angles to fit the existing data.

It is possible to make further simplication. From (\ref{eq:thetaC}) we can 
see that the physical Cabibbo angle is the difference between the up and
down quark mixing angles. In other words, physics is independent of 
basis \cite{Dicus:2010iq}. We can choose a basis in which $\theta^d_{12} = 0$ and
$\theta^\nu_{12} = 45^\circ + \theta^u_{12}$, reducing $G^d_{12}$ to
$d_1$ and the combination (\ref{eq:full0}) becomes,
\begin{equation}
  \{ G_{12}(45^\circ + \theta^u_{12}), d_1, G_{12}(\theta^u_{12}) \} G^m_{23} \,.
\label{eq:full2}
\end{equation}
By using (\ref{eq:dG12}) to remove $d_1$, the combination can be 
replaced by,
\begin{equation}
  \{ G_{12}(\pm [45^\circ + \theta^u_{12}]), G_{12}(\pm \theta^u_{12}) \} G^m_{23} d^n_1 \,,
\label{eq:full3}
\end{equation}
where $n = 0, 1$. Although there are four different $G^f_{12}$ elements,
only one degree of freedom, namely $\theta^u_{12}$, is involved. Implementing
(\ref{eq:G12product}) and (\ref{eq:dG12}) repeatedly, the first factor
in (\ref{eq:full3}) can be replaced with a single $G^f_{12}$ element,
\begin{equation}
  G_{12}(i \times 45^\circ) G_{12}(j \times \theta^u_{12}) G^m_{23} d^n_1 \,,
\label{eq:full4}
\end{equation}
where $i$ and $j$ are integers. As long as $\theta^u_{12}$ is a rational 
ratio of $2 \pi$, the unified group is finite.

In common practice, we should also include $d_1$ and $d_2$, which are needed 
to constrain the charged leptons within a diagonal basis, to 
reconstruct the full flavor symmetry. Nevertheless, we find it difficult
to prove that the generated group is finite or not, although incorporating
$d_1$ only is straightforward. So we decide to leave
it as it is. After all, the residual symmetry in the charged lepton sector
has many degrees of freedom. Further, the full symmetry needs not to be 
finite, not even discrete. The bottom-up approach can reconstruct
the minimal part but may not the full symmetry. 
As long as the residual symmetry can 
dictate the mixing patterns, the role of flavor symmetry 
is fulfilled perfectly. The case of $d_1 \times d_2$ we have 
been using in this paper is just a handy example to illustrate our main idea of 
generalizing and unifying the residual symmetries in both the lepton and
quark sectors with the benefit of explaining the quark-lepton complementarity.
Like we have pointed out the importance of residual symmetries in our earlier
works \cite{residual1,residual2}, without reconstructing the full flavor
symmetry, we hope the current work can provide a new direction to move 
forward.

\noindent {\bf Acknowledgements}--
This work is supported in part by Grant-in-Aid for Scientific research (No. 25400287) from JSPS.

\end{document}